\documentclass[]{article}
\usepackage{threeparttable}
\usepackage{lmodern}
\usepackage{amssymb,amsmath}
\usepackage{ifxetex,ifluatex}
\usepackage[numbers,sort&compress]{natbib}
\ifnum 0\ifxetex 1\fi\ifluatex 1\fi=0 
  \usepackage[T1]{fontenc}
  \usepackage[utf8]{inputenc}
\else 
  \ifxetex
    \usepackage{mathspec}
  \else
    \usepackage{fontspec}
  \fi
  \defaultfontfeatures{Ligatures=TeX,Scale=MatchLowercase}
\fi
\IfFileExists{upquote.sty}{\usepackage{upquote}}{}
\IfFileExists{microtype.sty}{%
\usepackage[]{microtype}
\UseMicrotypeSet[protrusion]{basicmath} 
}{}
\PassOptionsToPackage{hyphens}{url} 
\usepackage[unicode=true]{hyperref}
\hypersetup{
            pdfborder={0 0 0},
            breaklinks=true}
\usepackage{longtable,booktabs}
\IfFileExists{footnote.sty}{\usepackage{footnote}\makesavenoteenv{long table}}{}
\usepackage{graphicx,grffile}
\makeatletter
\def\maxwidth{\ifdim\Gin@nat@width>\linewidth\linewidth\else\Gin@nat@width\fi}
\def\maxheight{\ifdim\Gin@nat@height>\textheight\textheight\else\Gin@nat@height\fi}
\makeatother
\setkeys{Gin}{width=\maxwidth,height=\maxheight,keepaspectratio}
\IfFileExists{parskip.sty}{%
\usepackage{parskip}
}{
\setlength{\parindent}{0pt}
\setlength{\parskip}{6pt plus 2pt minus 1pt}
}
\ifx\paragraph\undefined\else
\let\oldparagraph\paragraph
\renewcommand{\paragraph}[1]{\oldparagraph{#1}\mbox{}}
\fi
\ifx\subparagraph\undefined\else
\let\oldsubparagraph\subparagraph
\renewcommand{\subparagraph}[1]{\oldsubparagraph{#1}\mbox{}}
\fi

\makeatletter
\def\fps@figure{htbp}

\makeatother

\date{}
\title{Computational prediction of ideal strength for a material}
\begin{document}
\author{Zixun Wang\textsuperscript{1}, Xingyu Wang\textsuperscript{2}, Xianqi
Song\textsuperscript{1}, Xinxin Zhang\textsuperscript{3}, \\Hanyu
Liu\textsuperscript{1,*}, 
Miao Zhang\textsuperscript{2,*}}

\maketitle

\textsuperscript{1}\emph{Key Laboratory of Material Simulation Methods
and Software of Ministry of Education \& State Key Laboratory of
Superhard Materials, College of Physics, Jilin University, Changchun
130012, China}

\emph{\textsuperscript{2}Department of Physics, School of Sciences,
Beihua University, Jilin 132013, China}

\textsuperscript{3}\emph{Department of Science, Shenyang University of
Chemical Technology, Shenyang 110142, China}

*Corresponding author: 

\emph{hanyuliu@jlu.edu.cn}(Hanyu Liu);

\emph{zhangmiaolmc@126.com}(Miao
Zhang)

\section{Abstract}

The ideal strength is crucial for predicting material behavior under
extreme conditions, which can provide insights into material limits,
guide design and engineer for enhanced performance and durability. In
this work, we present a method within an allows for the estimation of
tensile, shear, and indentation strengths in any crystallographic
direction or plane. We have examined the strain-stress relationships of
several well-known structures and compared our findings with previous
work, demonstrating the effectiveness of our approach. Moreover, we
performed extensive investigations into the indentation strength of
hexagonal WC, $\beta$-SiC, and MgAl\textsubscript{2}O\textsubscript{4}. The
current study uncovers the modes of structural deformation and the
underlying atomistic mechanisms. The insights gained from this study
have significant implications for the further exploration and design of
superhard materials.

\textbf{Keywords:} First-principles calculations; Ideal strength;
Tensile strength; Shear strength; Indentation strength.

\section{Introduction}

The ideal elasticity strength is the minimum stress to plastically
deform an infinite dislocation-free crystal \cite{1}, which gives an
upper bound on the strength of a perfect crystal. The ideal strength
could be classified as tensile, shear, and indentation strength
according to the type of external force posted on the material. It is an
important parameter to evaluate the shear and indentation strengths are
believed to be closely related to the Vickers hardness of an ideal
crystal. Therefore, it is of technical and scientific interest to
investigate the ideal elasticity strength of a crystal. In industrial
practice, stress-strain relationships are commonly employed to determine
ideal strength \cite{2}, where various strains are applied to the crystal
and the corresponding stress is measured. Theoretical simulation of the
ideal strength under complex loading conditions by using stress-strain
relationship can also provide fundamental insights into the evolution of
chemical bonding and physical property changes of a materials, which is
helpful for elucidating and predicting anomalous behaviors of material
at extreme conditions \cite{2,3,4,5}.

Great efforts had been dedicated to the accurate calculation of the
ideal strength since the mid of 1980s\cite{1}. Initially, some
experimentally proposed semi-empirical models were used. It is commonly
believed that the tensile and shear strength of materials are directly
proportional to their Vickers hardness. However, the results are always
inaccurate. Recent advances in computational physics made it possible to
ab-initially calculate the ideal strength, and the ideal elasticity
strength of tremendous materials has been precisely calculated within
the framework of electronic structure\cite{1,6,7,8,9,10,11,12,13,14,15,16,17,18,19,20,21,22,23,24,25,26,27,28,29}. Indentation is a
method used in experiments to measure the hardness and strength of
materials. A more complex calculation of indentation strength has been
performed in the framework of DFT \cite{16,27,28}. Energy and force can
be calculated faster and more accurately using the DFT approach.
Therefore, this calculation can be applied in more scenarios. The
calculations of the elastic strength have been extensively applied to an
increasing number of materials, solving a wide range of engineering and
physical problems \cite{1,6,7,8,9,10,11,12,13,14,15,16,17,18,19,20,21,22,23,24,25,26,27,28,29}. Li \emph{et al}. investigated
boron-rich tungsten borides, and demonstrated novel ultrahard properties
with stress-strain behaviors revealing diverse and anomalous patterns.
The profound influence of boron concentration on bonding configurations
and deformation modes in WB\textsubscript{n} (n = 2, 3, 4) results in
distinct stress responses and unexpected variations in indentation
strength, highlighting the unique deformation mechanisms and the need to
explore unconventional structure-property relations in ultrahard
materials.\cite{21}Lu et al. designed as new-generation superhard
materials, transition-metal light-element compounds have been hindered
by indentation strain softening, limiting their intrinsic hardness below
40 GPa. Moreover, the calculations revealed that hP4-WN and
hP6-WN\textsubscript{2} tungsten nitrides exhibit extraordinary strain
stiffening, leading to enhanced indentation strengths exceeding 40 GPa,
opening possibilities for nontraditional superhard materials.\cite{24}

In this work, we provided a comprehensive exposition of the underlying
principles of our newly proposed methodology on calculating the tensile,
shear and indentation strengths based on Vienna Ab initio Simulation
Package (VASP) code, accompanied by comparative analysis with other
approaches to establish its robustness. We selected hexagonal WC, $\beta$-SiC,
and MgAl\textsubscript{2}O\textsubscript{4} for an in-depth
investigation of ideal strength. Additionally, we examined the
indentation strength of hexagonal WC, $\beta$-SiC, and
MgAl\textsubscript{2}O\textsubscript{4}.This comprehensive exploration
has afforded us a deeper comprehension of their mechanical
characteristics.

\section{Computational Methods}

Mathematically, the stress tensors are the second-order symmetric tensor
with nine components (three positive stress componentsand six shear
stress components), in which only six of them are independent. They also
can be represented by a 3×3 matrix, where each matrix element
corresponds to a stress component. To calculate the stress-strain
relationship, it should strain the crystal in a series of incremental
simple tensile or shear, and then relax the lattice as a function of the
strain to obtain the total energy \emph{E}\textsubscript{tot}. According
to DFT, \emph{E}\textsubscript{tot} can be obtained by

\[{E_{tot}} = \sum\limits_i {\left. {\left\langle {{\psi _i}\left| {\left. { - \frac{1}{2}{\nabla ^2}} \right|{\psi _i}} \right.} \right.} \right\rangle }  + \int {{V_{ext}}} (r){d^3}r + {E_H}(\rho ) + {E_{xc}}(\rho ) + {U_{II}}\]

where \emph{$\Psi$\textsubscript{i}} represents wavefunction,
\emph{V\textsubscript{ext}} is external potential,
\emph{E\textsubscript{H}}(\emph{$\rho$}) is the Hartree energy,
\emph{E\textsubscript{xc}}(\emph{$\rho$}) is exchange and correlation energy,
and \emph{U\textsubscript{II}} is the ion-ion interaction. \emph{$\rho$(r)}
is the density of the electron gas, which can be calculated by

\[\rho (r){\text{ = 2}}{\sum\limits_i {\left| {{\psi _i}} \right|} ^2}\]

Based on Hellmann-Feynman force theorem, each stress component is the
derivative of the energy with respect to the strain

\[{\tau _{mn}} =  - \frac{1}{V}\frac{{\partial {E_{tot}}}}{{\partial {\varepsilon _{mn}}}} =  - \left. {\left\langle {{\psi _i}\left| {\left. {\frac{{\partial \widehat H}}{{\partial {\varepsilon _{mn}}}}} \right|{\psi _i}} \right.} \right.} \right\rangle  =  - \frac{{\partial {U_{II}}}}{{\partial {\varepsilon _{mn}}}} - \int {\frac{{\partial {V_{ext}}(r)}}{{\partial {\varepsilon _{mn}}}}\rho ({\varepsilon _{mn}}){d^3}{\varepsilon _{mn}}} \]

where H represents the Hamiltonian, and \emph{$\varepsilon$\textsubscript{mn}}
represents strain.

According to the above method, VASP can simply determine the six
independent stress tensors \emph{$\tau$\textsubscript{xx}},
\emph{$\tau$\textsubscript{yy}}, \emph{$\tau$\textsubscript{zz}},
\emph{$\tau$\textsubscript{xy}}, \emph{$\tau$\textsubscript{xz}}, \emph{and
$\tau$\textsubscript{yz}} by a structural optimizations, if a strain is
induced in a well optimized crystal lattice by modifying the lattice
parameters, which are represented by a lattice matrix in VASP code. For
instance, if calculating the tensile stress along {[}100{]} direction in
fcc lattice, one just needs to add the tensile strain along the
{[}100{]} direction (or x-axis of crystal lattice) by changing the
corresponding lattice matrix element, and then optimize the structure by
fixing the \emph{F}\textsubscript{xx} component of the optimization
matrix \emph{F}\textsubscript{mn} (m, n = x, y, z) in VASP code without
optimization. Then we can obtain the tensile stress of
\emph{$\tau$\textsubscript{xx}} which corresponds to the studied tensile
stress \emph{$\tau$}\textsubscript{{[}100{]}}. Considering that VASP can
straightforwardly compute the stress tensor of
\emph{$\tau$\textsubscript{xx}}, \emph{$\tau$\textsubscript{yy}},
\emph{$\tau$\textsubscript{zz}}, if the tensile strain was added to the
coordinate axis of crystal lattice. Therefore, we rotate the studied
tensile direction {[}\emph{uvw}{]} to one of coordinate axis (here, we
choose \emph{x}-axis). In this way, after structural optimization by
setting \emph{F}\textsubscript{xx} to zero, we can obtained the tensile
stress of \emph{$\tau$\textsubscript{xx }}which corresponds to the studied
tensile stress \emph{$\tau$}\textsubscript{{[}\emph{uvw}{]}}. By gradually
applying a series of continuous strain along the tensile direction and
ultimately deriving the corresponding tensile stress, we can obtain the
tensile stress-strain relationship.

It is obvious that during the tensile strength calculations, one
important step is to rotate the studied tensile direction to the
\emph{x}-axis. Here, take the orthorhombic lattice as an example to
introduce how to rotate the studied tensile direction to \emph{x}-axis.
Using Cartesian coordinate system, the actual stretching direction is OA
as shown in Fig. 1(a). It can be achieved by rotating -$\alpha$~degree (here
counterclockwise rotation is set to be positive direction) around the
\emph{z}-axis (see Fig. 1b), then rotate $\beta$ degree around the
\emph{y}-axis.
\begin{figure}[ht]
\centering
\includegraphics[width=4.53403in,height=1.92153in]{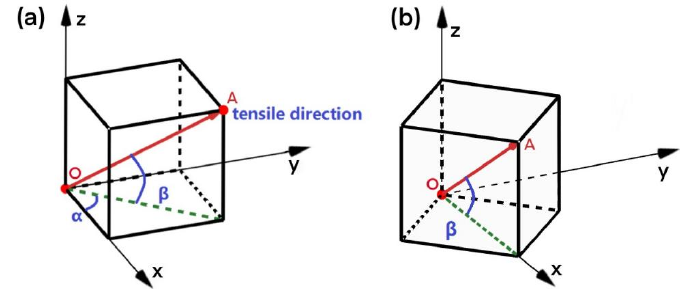}

\textbf{Fig. 1} Rotation of tensile direction in an orthohormbic
lattice. (a) tensile direction before rotation. The red vector denotes
the studied tensile direction, the green line is its projection on
\emph{xy}-plane, $\alpha$ is the angle of projection to the x-axis, $\beta$ is the
angle of tensile direction to xy-plane. (b) The lattice after tensile
direction rotating -$\alpha$ around \emph{z}-axis, in which the tensile
direction is located in \emph{xz}-plane.
\end{figure}
Shear strength is more complicated than tensile strength. Shear stress
calculations need to rotate the studied shear direction {[}\emph{uvw}{]}
to z-axis and reorient the normal vector of the shear plane to align
with \emph{x}-axis. Consequently, by maintaining F\textsubscript{xz} as
zero during structural optimization, we can calculate the shear stress
\emph{$\tau$\textsubscript{xz},} which is corresponding to the studied shear
stress \emph{$\tau$}\textsubscript{{[}\emph{uvw}{]}}. Here, to give a more
vivid illustration on the rotation process of shear plane and shear
direction, orthorhombic lattice was also taken as an example. As it is
can be seen in Fig. 2(a), the sheared plane is characterized by its
normal vector OA and the shear direction OB lies within this plane.
Firstly, the normal vector OA was rotated -$\alpha$ degree around \emph{z}-axis
{[}see Fig. 2(b){]}, and then $\beta$ degree around \emph{y}-axis {[}see Fig.
2(c){]}. Subsequently, rotate OA $\gamma$ degree around the x-axis{[}Fig.
2(d){]}. Finally, the normal vector is align with the x-axis, and the
shear stress direction aligns with the z-axis.
\begin{figure}[ht]
\centering
\includegraphics[width=5.17153in,height=3.64722in]{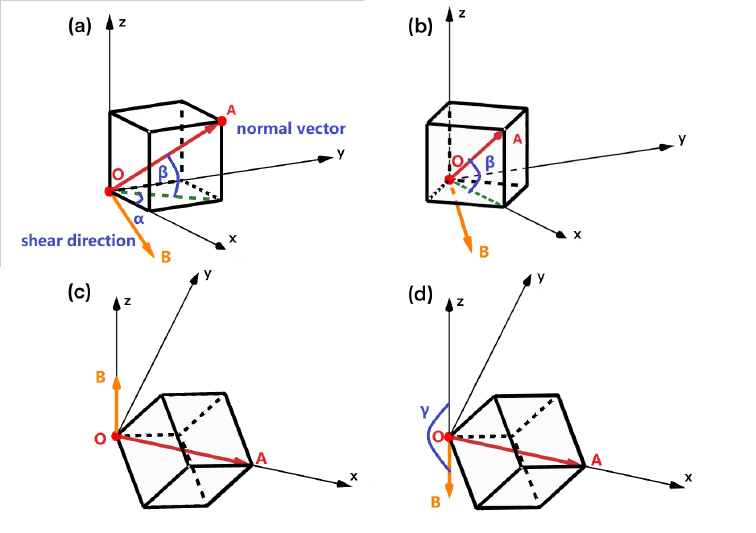}

\textbf{Fig. 2} Rotation of shear plane and direction in an orthohormbic
lattice. (a) The orthorhombic lattice with shear plane and direction
marked out. The yellow vector represents shear direction, and the red
vector represents normal vector of the shear plane. The green line is
the projection on xy-plane. $\alpha$ is the angle of projection to the x-axis,
and $\beta$ is the angle of tensile direction to xy-plane. (b) The lattice
after rotating -$\alpha$ around z-axis to make normal direction in xz-plane.
(c) The lattice after rotating $\beta$ around y-axis to make tensile direction
in x-axis. (d) The lattice after rotating $\gamma$ around y-axis to make shear
direction in z-axis.
\end{figure}
The indentation strength is a more complex case, where in actual
compression the material receives not only tensile stress perpendicular
to the plane, but also in-plane shear stress. Experiments have shown
that there is a close relationship between nano-scale indentation and
the geometry of the indenter \cite{16,22,23,29}. When using the
first-principles method to calculate compression, we assume that the
material is uniformly deformed within a small volume. And the
deformation is related to the shape of the indenter. Such a uniformly
deformed lattice is sufficient in the study of the breakage of
interactions on the atomic microscopic scale. In the first-principles
calculations, we can simplify the situation to a uniform deformed
infinite and perfect lattice under a uniform stress field\cite{30}. We
assume that a sharp indenter squeezes the material by the way shown in
the Fig. 3(a). The tensile stress $\tau$\emph{\textsubscript{xx}} and shear
stress $\tau$\emph{\textsubscript{xz}} received by the material under a sharp
indenter always maintain the following relationship:
$\tau$\emph{\textsubscript{xz}} =$\tau$\emph{\textsubscript{xx}}tan$\theta$, where $\theta$ is
determined by the shape of the indenter, as shown in Fig. 3(b).
\begin{figure}[ht]
\centering
\includegraphics[width=4.5in,height=1.45903in]{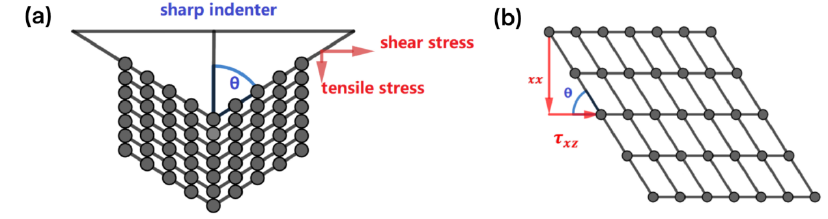}

\textbf{Fig. 3} Microscopic indentation. (a) A sharp indentation acts on
a material, and 2$\theta$ is the angle of indenter. (b) The microscopic lattice
beneath the indenter.
\end{figure}
Based on this relationship, we only need to calculate the shear stress
by using the method of calculating shear stress proposed in this work.
Note that since we consider the crystal is an infinitely large perfect
lattice with uniform deformation, we must keep the angle of the
deformation constant in both directions. This can be achieved by
modifying the optimization program in VASP code. For a blunt indenter,
as long as we know that the direction of the tangent at each point of
contact surface between the indenter and the material, we can use the
approach of a sharp indenter to calculate the indentation stress at each
point of the blunt indenter, as shown in Fig. 4\cite{30}.
\begin{figure}[ht]
\centering
\includegraphics[width=3in,height=3.13819in]{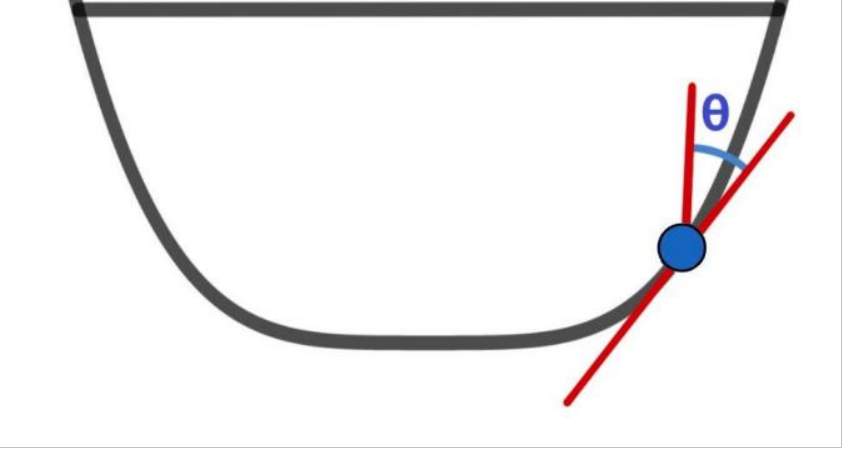}

\textbf{Fig. 4} Blunt indenter.
\end{figure}
\section{Results and discussion}

To thoroughly validate the accuracy and reliability of our approach, we
conducted an extensive analysis involving diamond, Si, hexagonal AlN,
and $\alpha$-Al\textsubscript{2}O\textsubscript{3}. The tensile, shear and indentation strengths were calculated separately by setting the relative strain at each step to 0.01 times the elements of the lattice matrix. We used the VASP package with the Perdew-Burke-Ernzerhof (PBE) version of the generalized gradient approximation (GGA) for the exchange-correlation functional and a plane-wave basis set \cite{31,32,33,34}. For the integration in the Brillouin zone, we set the k-points to 6 × 6 × 6 for all the calculated material. The cutoff energy is set to 600 eV, which is sufficient to make the total energy converge to an accuracy of less than 0.001 eV/atom.   The paramount objective of our investigation is to establish the veracity of our proposed
methodology. In pursuit of this goal, we performed a meticulous
comparison of our computed results against those obtained using
established methods. This validation process was meticulously conducted
under conditions consistent with prior research, ensuring a fair and
direct comparison.

The crystallographic configurations of the chosen materials are
presented in Table S1, while the outcomes of the tensile, shear, and
indentation are elaborated in Tables 1, respectively. The significance
of these tables lies in their role as a tangible demonstration of the
effectiveness of our method across various loading scenarios. In the
case of diamond, our calculations yielded the weakest tensile and shear
directions consistent with previous research. The weakest tensile
direction is {[}111{]}, and the weakest shear direction is {[}111{]}
\textless{}\(11\overline{2}\)\textgreater{}. Additionally, our Vickers
hardness value is 95.7 GPa, which closely aligns with the prior result
of 97.6 GPa, showcasing minimal deviation. This conforms to the
characteristics of diamond as a superhard material. For cases Si, the
weakest tensile and shear directions are {[}111{]} and {[}111{]}
\textless{}\(11\overline{2}\)\textgreater{}, respectively. And its
indentation strength is 6.1 GPa. Additionally, the weakest tensile
direction for hexagonal AlN is {[}\(\overline{1}2\overline{1}0\){]}, and
the weakest shear direction is {[}\(\overline{1}2\overline{1}0\){]}
\textless{}\(10\overline{1}0\)\textgreater{}. The indentation strength
is 11.7 GPa. Moreover, the weakest tensile and shear directions are
{[}\(10\overline{1}0\){]} and {[}\(10\overline{1}0\){]}
\textless{}0001\textgreater{} for
$\alpha$-Al\textsubscript{2}O\textsubscript{3}. The indentation strength is
28.9 GPa. Our computed results also closely match those from the
previous work. The specific stress-strain curves for tension, shear, and
indentation of these four materials are presented in Figures S1-4,
respectively.

\begin{table}[]
\caption{Ideal tensile, shear and indentation strength in three
directions for diamond, Si, hex-AlN, and
\emph{$\alpha$}-Al\textsubscript{2}O\textsubscript{3}, where hex denotes
hexagonal phase.}
\resizebox{\linewidth}{!}{
\begin{threeparttable}
\begin{tabular}{cccccccccc}
\hline
\multicolumn{1}{l}{\textbf{Structure}} & \multicolumn{3}{c}{\textbf{Tensile strength (GPa)}} & \multicolumn{3}{c}{\textbf{Shear strength (GPa)}}                                        & \multicolumn{3}{c}{\textbf{Indentation strength (GPa)}}                                                 \\ \hline
                                       & Direction           & This work       & Other       & Direction                                                            & This work & Other & Direction                                              & This work             & Other                  \\
{diamond}               & {[}100{]}           & 225.7           &             & {[}100{]} \textless{}001\textgreater{}                               & 140.4     &       & {[}111{]} \textless{}11$\overline{2}$\textgreater{} & 95.7 & 97.6$^e$ \\
                                       & {[}110{]}           & 126.4           &             & {[}110{]} \textless{}001\textgreater{}                               & 98.5      &       &                                                        &                       &                        \\
                                       & {[}111{]}           & 92.4            & 95$^a$         & {[}111{]} \textless{}11$\overline{2}$\textgreater{} & 93.7      & 93.0$^a$  &                                                        &                       &                        \\
{Si}                    & {[}100{]}           & 86.8            &             & {[}100{]} \textless{}001\textgreater{}                               & 12.1      &       & {[}111{]} \textless{}11$\overline{2}$\textgreater{} & 6.1  &       \\
                                       & {[}110{]}           & 29.5            &             & {[}001{]} \textless{}100\textgreater{}                               & 15.0      &       &                                                        &                       &                        \\
                                       & {[}111{]}           & 21.1            & 22$^b$         & {[}110{]} \textless{}001\textgreater{}                               & 7.5       & 6.8$^b$  &                                                        &                       &                        \\
{Hex-AlN}               & {[}0001{]}          & 38.1            & 39c         & {[}10$\overline1$0{]} \textless{}0001\textgreater{}                            & 20.2      & 20.0$^c$ &{[}10$\overline1$0{]} \textless{}$\overline1$2$\overline1$0\textgreater{}                & 11.7                  &                        \\
                                       & {[}10$\overline1$0{]}         & 38.0            & 38$^c$         & {[}$\overline1$2$\overline1$0{]} \textless{}0001\textgreater{}                           & 20.0      & 22.5$^c$ &                                                        &                       &                        \\
                                       & {[}$\overline1$2$\overline1$0{]}        & 32.1            & 34$^c$        & {[}10$\overline1$0{]} \textless{}$\overline1$2$\overline1$0\textgreater{}                                                          & 19.6      & 19.9$^c$ &                                                        &                       &                        \\
\multicolumn{1}{l}{}                   & {[}0001{]}          & 57.3            & 58.3$^d$       & {[}10$\overline1$0{]} \textless{}0001\textgreater{}                                   & 16.4      &       &                                                        &                       &                        \\
\textit{$\alpha$-Al2O3}                       & {[}10$\overline1$0{]}         & 26.3            & 26.3$^d$       & {[}$\overline1$2$\overline1$0{]} \textless{}0001\textgreater{}                                & 14.5      & 14.4$^d$ & {[}$\overline1$2$\overline1$0{]} \textless{}0001\textgreater{}                      & 28.9                  &                        \\
\multicolumn{1}{l}{}                   & {[}$\overline1$2$\overline1$0{]}        & 36.5            & 36.5$^d$       & {[}10$\overline1$0{]} \textless{}$\overline1$2$\overline1$0\textgreater{}                                & 18.7      &       &                                                        &                       &                        \\ \hline
\end{tabular}
\begin{tablenotes}
        \footnotesize
        \item a Ref \cite{12} b Ref \cite{18} c Ref \cite{10} d Ref \cite{11} e Ref \cite{16}
      \end{tablenotes}
\end{threeparttable}
}
\end{table}
We subsequently placed particular emphasis on an in-depth investigation
of the ideal strengths of three distinct materials: WC, $\beta$-SiC,and
MgAl\textsubscript{2}O\textsubscript{4}. We would like to calculate the
Vickers hardness, so the indenter we consider to be a Vickers indenter
with an angle of 136°between the opposing surfaces.

Hexagonal tungsten carbide (WC) is crucial for cutting-edge industries
like manufacturing, aerospace, and electronics, thanks to its
exceptional properties. Its unique combination of hardness, strength,
and thermal stability significantly enhances overall performance and
efficiency. We extensively undertook an investigation into the ideal
tensile strength, ideal shear strength, and indentation strength of
hexagonal tungsten carbide (WC). According to academic standards,
Fig.5(a) illustrates several distinctive crystallographic orientations
of the WC crystal, which predominantly guided our calculations in this
study. We conducted tensile calculations to explore these properties
along the {[}0001{]}, {[}\(10\overline{1}0\){]} and
{[}\(\overline{1}2\overline{1}0\rbrack\) crystallographic directions.
The tensile strengths were found to be 101.3 GPa, 72.6 GPa, and 63.5 GPa
in the {[}0001{]}, {[}\(10\overline{1}0\){]} and
{[}\(\overline{1}2\overline{1}0\rbrack\) orientation, respectively.
Fig.5(b) visually presents the stress-strain curves for the investigated
directions, highlighting {[}\(\overline{1}2\overline{1}0\rbrack\) as the
least resistant tensile orientation within WC.

We next undertook an in-depth examination involving shear calculations
of properties along the
{[}\(\overline{1}2\overline{1}0\rbrack\)\textless{}0001\textgreater{},
{[}\(10\overline{1}0\){]}\textless{}0001\textgreater{}, and
{[}\(\overline{1}2\overline{1}0\){]}
\textless{}\(10\overline{1}0\)\textgreater{} crystallographic
orientations. The shear strength along the
{[}\(\overline{1}2\overline{1}0\rbrack\)\textless{}0001\textgreater{}
direction emerged at an impressive 41.6 GPa, while the
{[}\(10\overline{1}0\){]}\textless{}0001\textgreater{} orientation
showcased a noteworthy tensile strength of 33.9 GPa. Additionally, the
{[}\(\overline{1}2\overline{1}0\){]}
\textless{}\(10\overline{1}0\)\textgreater{} orientation unveiled a
distinctive tensile strength of 18.7 GPa. These significant findings are
shown in Fig.5(c), elegantly capturing the essence of stress-strain
profiles for each respective orientation. Remarkably,
{[}\(\overline{1}2\overline{1}0\){]}
\textless{}\(10\overline{1}0\)\textgreater{} represents the weakest
shear direction of WC. Therefore, we selected the weakest shear
direction as the orientation for its indentation.

A comprehensive analysis of the indentation process was conducted and
visualized in Fig.5(d). Intriguingly, as the strain surpassed 0.27 times
the crystal length in the aforementioned direction, the chemical bonds
between tungsten and carbon atoms suffered rupture due to increased
atomic separation. At this point, the indentation strength of WC is
36.8GPa. Its stress-strain curve is displayed in Fig.5 (e). This
phenomenon is attributed to the exceeding distance between the atoms,
leading to bond fracture. Our study provides valuable insights into the
mechanical response of hexagonal WC under various loading conditions and
crystallographic orientations, shedding light on the intricate interplay
between atomic bonds and mechanical deformation. The calculation results
indicate that WC is a relatively hard material with the potential to
become a superhard material.
\begin{figure}[ht]
\centering
\includegraphics[width=4.5in,height=3.13819in]{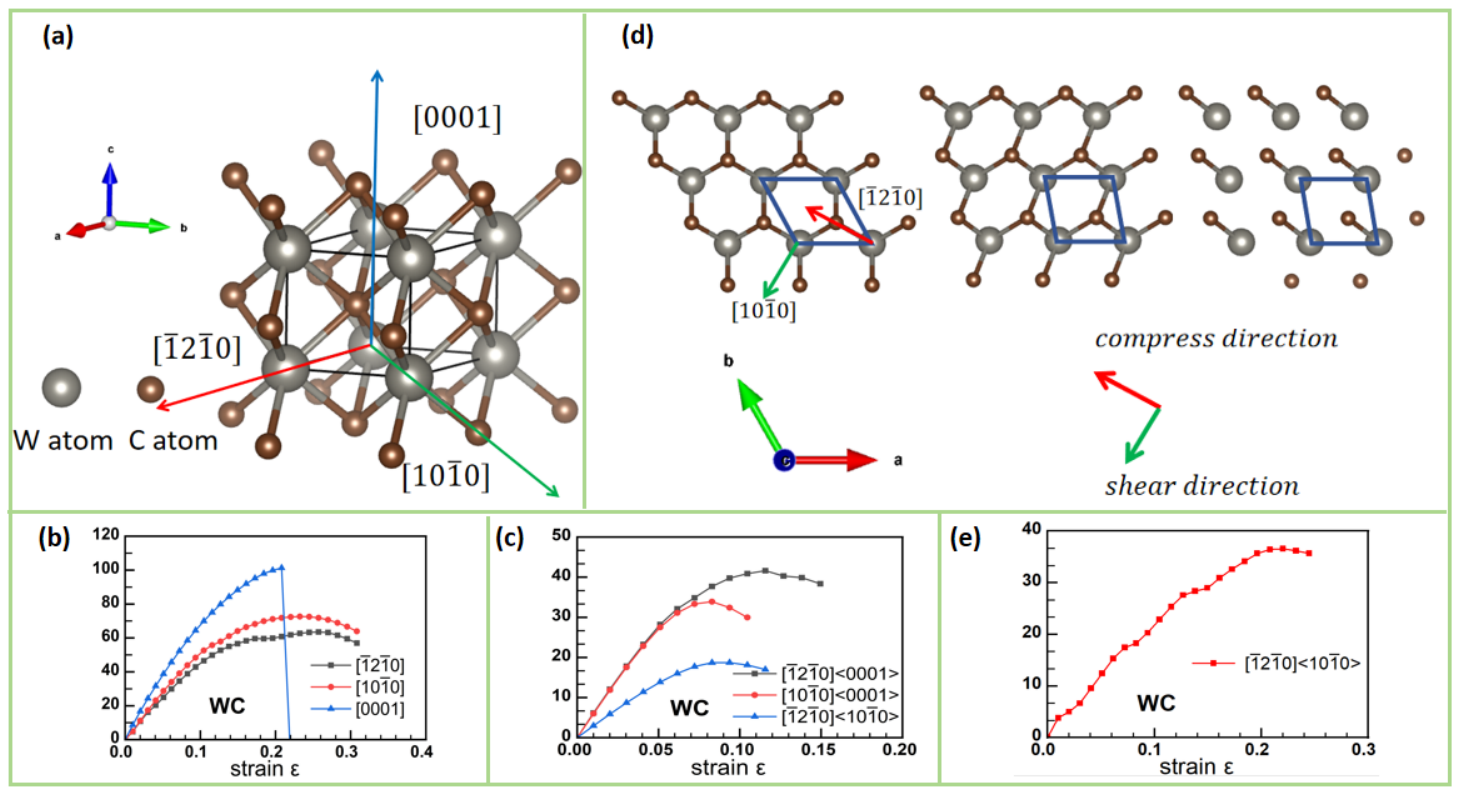}

\textbf{Fig. 5} (a)Structure and the crystallographic directions of WC
(b) The tensile stress-strain curve. (c)The shear stress-strain curve.
(d) Structural changes during the indentation process and fracture of
chemical bonds. (e)The indentation stress-strain curve.
\end{figure}
At equilibrium condition,
\protect\hypertarget{_Hlk143064073}{}{}\emph{$\beta$}-SiC adopts the cubic
structure in \emph{F}-43\emph{m} symmetry {[}Fig. 6(a){]}, where each Si
atom is connected to four C atoms, forming a strong
\emph{sp}\textsuperscript{3} covalent bond unit with Si-C bond lengths
of 1.89 Å. We first set out to determine the easy cleavage plane of
\protect\hypertarget{_Hlk143064676}{}{}\emph{$\beta$}-SiC by simulating its
ideal tensile strength along several high-symmetry directions. The
calculated results shown in Fig. 6(b) clearly indicate that the weakest
peak tensile stress occurs in the {[}111{]} direction in
\protect\hypertarget{_Hlk143064194}{}{}\emph{$\beta$}-SiC, giving rise the
(111) easy cleavage planes. The result is 45 GPa in
\textless{}111\textgreater{} direction for \emph{$\beta$}-SiC and the Si-C
bonds soften considerably under the tensile strain long before the bond
breaking. Since a material's hardness is primarily determined by its
ability to resist shear deformation, we examine shear stress responses
of \emph{$\beta$}-SiC in its major low-index planes to establish key
benchmarks. We show in Fig.6(c) the stress responses under pure and
indentation shear strains along the
\protect\hypertarget{_Hlk143066266}{}{}(111)
\protect\hypertarget{_Hlk142285581}{}{}{[}11\(\overset{\overline{}}{2}\){]}
direction. The two sets of shear stresses reach peak values 29.7 and
19.7 GPa, respectively. At the indentation strain
(\protect\hypertarget{_Hlk143066722}{}{\protect\hypertarget{_Hlk143066028}{}{}}$\varepsilon$
= 0.36), the stress has only decreased from the peak value. The peak
pure shear stress occurs at the strain $\varepsilon$ = 0.22 but the interlayer
\emph{sp}\textsuperscript{3} bonding persists up to $\varepsilon$ = 0.32 where
significant buckling still exists and the Si-C bond length increased to
2.076 Å. The corresponding structural snapshot shows significant
buckling in the (111) planes, indicating a strong
\emph{sp}\textsuperscript{3} bonding character. The structural failure
occurs at $\varepsilon$ = 0.38, where the interlayer bonds in the (111)
{[}11\(\overset{\overline{}}{2}\){]} direction break up and the layers
become essentially flat, signaling a dominant
\emph{sp}\textsuperscript{2} bonding character. Here we notice that all
the Si-C bonds responsible for the structural failure also break
simultaneously under shear strains.The calculation results show that
\emph{$\beta$}-SiC possesses favorable mechanical properties.
\begin{figure}[ht]
\centering
\includegraphics[width=4.5in,height=2.75556in]{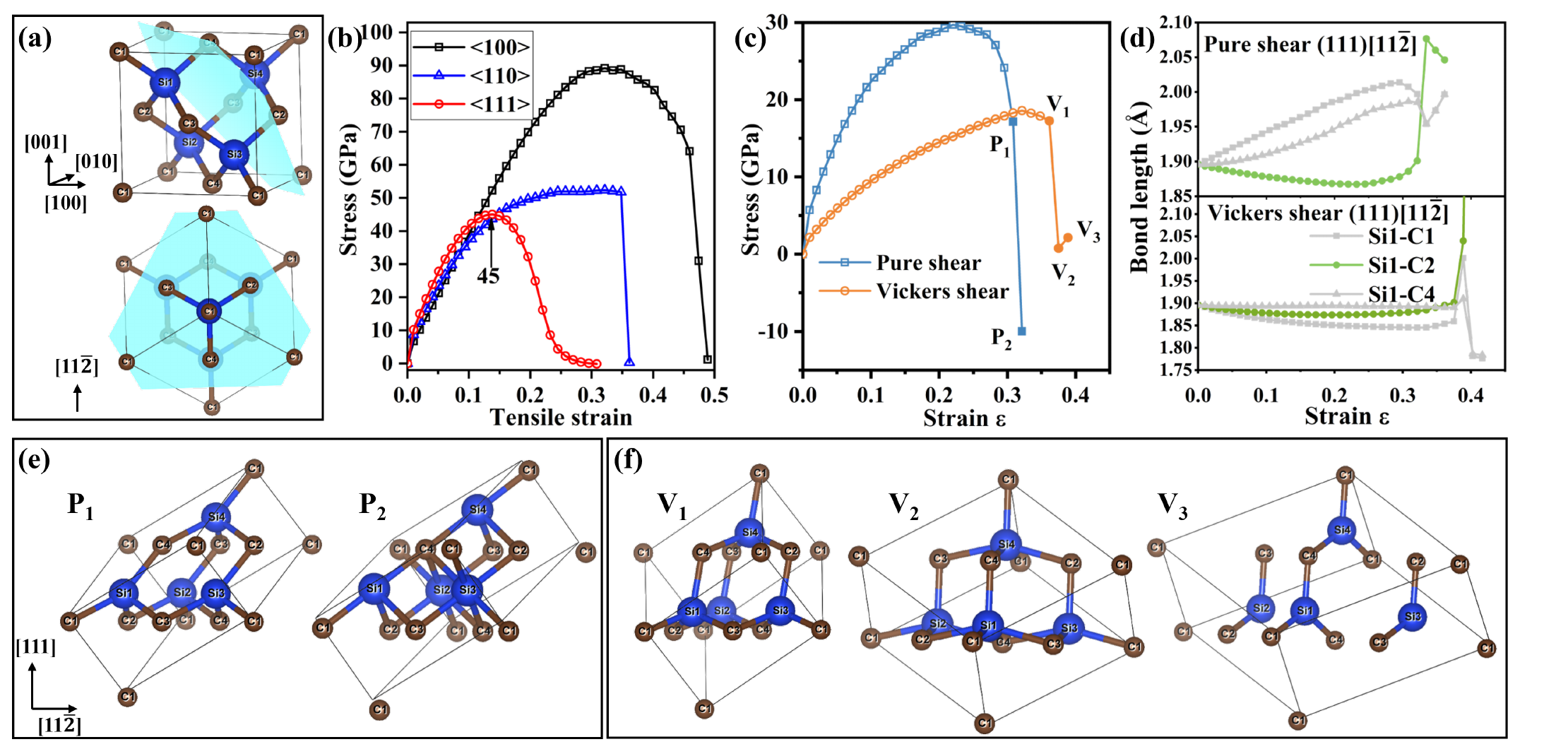}

\textbf{FIG.6.} (a) The crystal structure of
\protect\hypertarget{_Hlk143062749}{}{}\emph{$\beta$}-SiC at equilibrium (top)
and projected onto the (111) plane (bottom). (b) Calculated stress
responses under tensile strains along various high-symmetry directions
of \emph{$\beta$}-SiC. (c) Calculated stress responses under pure shear and
indentation shear strains in the (111) plane along the
\protect\hypertarget{_Hlk143062964}{}{}{[}11\(\overset{\overline{}}{2}\){]}
shear direction. (d) Bond-length changes under the (111)
{[}11\(\overset{\overline{}}{2}\){]} pure shear and indentation strains.
Corresponding snapshots of selected bonding configurations at key
deformation points before and after the large drop of stress on each
stress-strain curve under \protect\hypertarget{_Hlk143063012}{}{}pure
shear and indentation strains are presented in (e) and (f),
respectively.
\end{figure}
MgAl\textsubscript{2}O\textsubscript{4} holds significant importance
across diverse fields due to its remarkable combination of properties.
As a transparent ceramic, it finds applications in optical systems,
lasers, and transparent armor. Its high-temperature stability,
electrical insulating properties, and chemical resistance make it
valuable in aerospace, electronics, and corrosion-resistant coatings.
Moreover, spinel's biocompatibility has fueled its adoption in medical
implants. Its versatile attributes and extensive utility underscore
spinel's role as a multifunctional material with widespread impact.

Our investigation centered around the characterization of spinel
MgAl\textsubscript{2}O\textsubscript{4} in terms of its ideal tensile
strength, ideal shear strength, and indentation strength. To adhere to
academic conventions, we adhered to Fig.7(a), which visually illustrates
distinct crystallographic orientations within the
MgAl\textsubscript{2}O\textsubscript{4} crystal. These orientations
served as the foundation for our computations in this study. Tensile
calculations were carried out along the {[}100{]}, {[}110{]}, and
{[}111{]} crystallographic directions, yielding corresponding strengths
of 89.2 GPa, 28.28 GPa, and 28.3 GPa, respectively. The stress-strain
curves for these orientations are depicted in Fig.7(b), with the
{[}111{]} orientation revealing the least resistance to tensile forces
within MgAl\textsubscript{2}O\textsubscript{4}.

Our subsequent focus was on an exhaustive exploration of shear
calculations along the
\(\lbrack 100\rbrack\)\textless{}001\textgreater{},
\(\lbrack 110\){]}\textless{}001\textgreater{}, and {[}111{]}
\textless{}\(11\overline{2}\)\textgreater{} crystallographic
orientations. Notably, the shear strength along
the\(\lbrack 100\rbrack\)\textless{}001\textgreater{} direction
registered an impressive 29.7 GPa, while the
\(\lbrack 110\){]}\textless{}001\textgreater{} orientation displayed a
remarkable tensile strength of 31.0 GPa. Furthermore, the {[}111{]}
\textless{}\(11\overline{2}\)\textgreater{} orientation exhibited a
distinctive tensile strength of 14.2 GPa. Fig6 (c) elegantly
encapsulates these pivotal findings, showcasing stress-strain profiles
for each orientation. Interestingly, the {[}111{]}
\textless{}\(11\overline{2}\)\textgreater{} orientation emerged as the
weakest shear direction within MgAl\textsubscript{2}O\textsubscript{4},
thereby being chosen for indentation analysis.

Our comprehensive analysis of the indentation process is vividly
presented in Fig. 7(d). Intriguingly, when the strain surpassed 0.46
times the crystal length in the aforementioned direction, the increased
atomic separation led to the rupture of chemical bonds. At this
juncture, the indentation strength of
MgAl\textsubscript{2}O\textsubscript{4} was determined to be 22.2 GPa,
as depicted in the stress-strain curve presented in Fig.7(e). This
phenomenon was attributed to the exceeding atomic distance, which
induced bond fracture. Our study thus yields valuable insights into the
mechanical behavior of hexagonal MgAl\textsubscript{2}O\textsubscript{4}
under varied loading conditions and crystallographic orientations,
unraveling the intricate interplay between atomic bonds and mechanical
deformation. Its hardness enables it to be applied in more diverse
applications.
\begin{figure}[ht]
\centering
\includegraphics[width=4.5in,height=3.11944in]{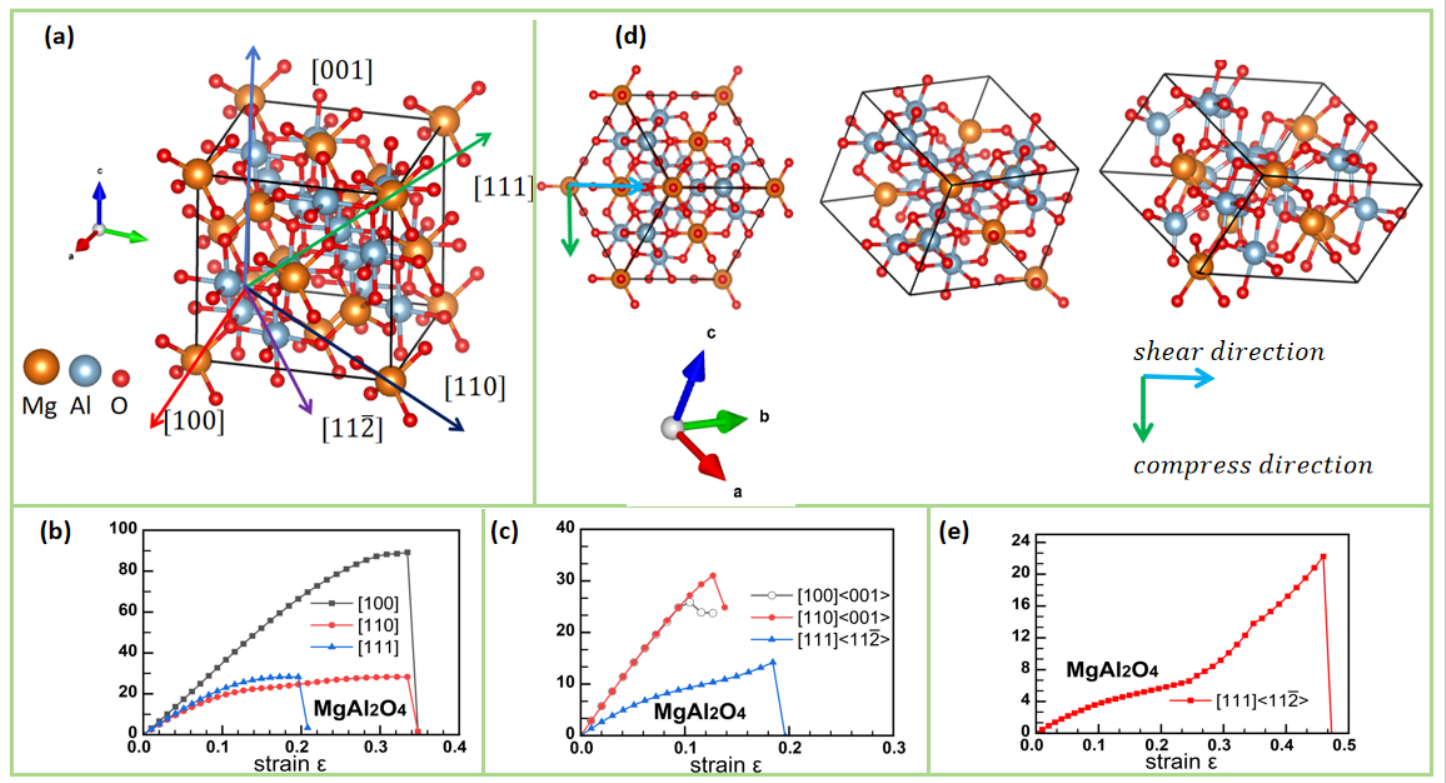}

\textbf{Fig. 7} (a)Structure and the crystallographic directions of
MgAl\textsubscript{2}O\textsubscript{4} (b) The tensile stress-strain
curve. (c)The shear stress-strain curve. (d) Structural changes during
the indentation process and fracture of chemical bonds. (e)The
indentation stress-strain curve.
\end{figure}
\section{Conclusions}

In summary, we introduced a novel methodology for computing the ideal
strength of materials based on first-principles calculations. The
approach enabled the determination of tensile, shear, and indentation
strengths in arbitrary crystallographic directions, facilitating a
comprehensive understanding of material response under different loading
conditions. Through a meticulous comparison with prior works, we
demonstrated the accuracy and reliability of our proposed method.
Furthermore, our investigation delved into the specific cases of
hexagonal WC, cubic SiC, and spinel
MgAl\textsubscript{2}O\textsubscript{4}, shedding light on their ideal
strengths and the structural transformations that occur under applied
stress. Based on our computational results, the Vickers hardness of WC
reaches 36.8 GPa, which is closed to 40 GPa. Therefore, it could not be
classified as a superhard material. While WC has the potential to become
a superhard material through various methods, such as
nanocrystallization, high-pressure synthesis, and alloying. Meanwhile,
cubic SiC and MgAl\textsubscript{2}O\textsubscript{4} exhibit Vickers
hardness within the range of 20-30 GPa, rendering them suitable for
diverse applications in various fields. This detailed analysis not only
enriched our understanding of these materials' mechanical properties but
also highlighted the versatility of our methodology in exploring a
diverse range of crystalline structures. The findings presented in this
study have broad implications for material design and engineering,
offering a rigorous foundation for predicting and optimizing the
mechanical behaviors of various materials. As we continue to refine and
expand our computational techniques, the insights gained from this
research will contribute to advancements across multiple disciplines,
driving innovation in materials science and engineering.
\section{Acknowledgments}
This work is supported by the Natural Science Foundation of China (Grant
No.12074138), and the Jilin Provincial Science and Technology
Development Joint Fund Project (Grant No. YDZJ202201ZYTS581).

\bibliographystyle{unsrt}
\bibliography{main}

\end{document}